\documentclass{mem}
\usepackage{natbib}\usepackage{txfonts}\usepackage{balance}
\usepackage{graphicx}
\usepackage[a4paper]{hyperref}
\idline{75}{282}
\begin{document}
\def\teff{$T\rm_{eff }$}
\def\kms{$\mathrm {km s}^{-1}$}

\title{The 2006 Radio Flare in the Jet of CTA\,102}

\author{C. M. Fromm\inst{1}, E. Ros\inst{2,1}, T. Savolainen\inst{1}, M. Perucho\inst{2}, A. P. Lobanov\inst{1}
          \and
          J. A. Zensus\inst{1}
          }
  \offprints{C. M. Fromm}
  
\institute{Max-Planck-Institut f\"ur Radioastronomie, Auf dem H\"ugel 69, D-53121 Bonn, Germany 
         \and
             Departament d'Astronomia i Astrof\'{\i}sica, Universitat de Val\`encia, E-46100, Burjassot, Val\`encia, Spain
\email{cfromm@mpifr.de}
}

\authorrunning{Fromm}

\titlerunning{}

\abstract{
The blazar CTA\,102 underwent a major radio flare in April 2006. We used several $15\,\mathrm{GHz}$ VLBI observations from the MOJAVE program to investigate the influence of this extreme event on jet kinematics. The result of modeling and analysis lead to the suggestion of an interaction between traveling and standing shocks $0.2\,\mathrm{mas}$ away from the VLBI core.  
}
\maketitle{}

\section{Introduction}
The radio source CTA\,102 ($z=1.037$) is one of the most observed AGN in the northern sky. Its observational history started in the late 1950s \citep{Har60}. Flux density variations were reported in the source \citep{Sho65}, which lead other authors to suggest that the signal was coming from an extraterrestrial civilization \citep{Kar64}. Later, CTA\,102 was identified as a quasar. Due to its strong flux density variability CTA\,102 has been the target for numerous observations at different wavelengths. 
\newline On kpc-scales CTA\,102 consists of a central core and two faint lobes \citep{Spe89}. The brighter lobe has a flux density of $170\,\mathrm{mJy}$ at a distance of $1.6\,\mathrm{arcsec}$ from the core at position angle (P. A.) of $143^\circ$ (measured from North through East). The other lobe, with a flux density of $75\,\mathrm{mJy}$, is located $1\,\mathrm{arcsec}$ from the center at P. A. $-43^\circ$. The spectral indices of the lobes are $-0.7$ for the brighter one and $-0.3$ for the other. 
\newline High resolution $15\,\mathrm{GHz}$ VLBI images show a curved jet with components which exhibit apparent velocities up to $15.4\pm 0.9\,\mathrm{c}$, adopting a cosmology with $\Omega_m=0.27$, $\Omega_\Lambda=0.73$ and $H_0=71\,\mathrm{km\,s^{-1}\, Mpc^{-1}}$. \citep{Lis09b}. \citet{Jor05} and \citet{Hov09} derived bulk Lorentz factors, $\Gamma$, between $15$ and $17$ and Doppler factors, $\delta$, between $15$ and $22$ from $43\,\mathrm{GHz}$ VLBI observations and from single-dish light curves.
\newline A major flux density outburst occurred in April 2006, which offers a unique opportunity to study CTA\,102 under these conditions \citep{Fro10}. For our analysis we used data from the monitoring of CTA\,102, performed within the framework of the $15\,\mathrm{GHz}$ MOJAVE\footnote{http://www.physics.purdue.edu/MOJAVE} program (Monitoring of Jets in Active galactic nuclei with VLBA Experiments; \citet{Lis09b}).

\section{Observations}
\begin{figure}[h!]
\resizebox{\hsize}{!}{\includegraphics[clip=true]{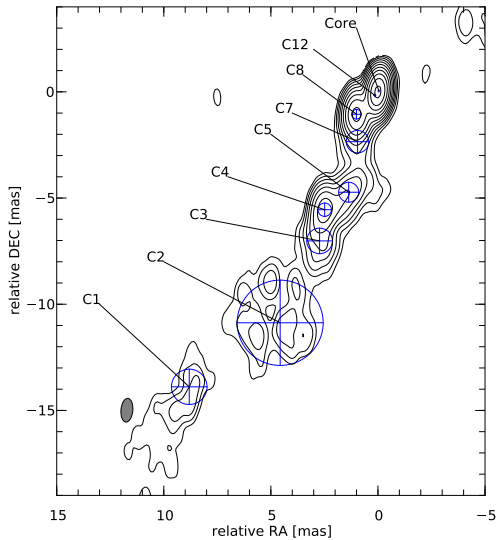}}
\caption{\footnotesize $15\,\mathrm{GHz}$ VLBI image of CTA\,102 with fitted circular Gaussian components observed on 6th of January 2007. The map peak flux is $1.13\,\mathrm{Jy/beam}$, where the convolving beam is $1.1\times 0.6\,\mathrm{mas}$ at P. A. $-4.8^\circ$.}
\label{VLBI}
\end{figure}

The MOJAVE observations of CTA\,102 between 2005 and 2008 have been used to analyze kinematic changes in the source during the 2006 radio flare. The raw data were calibrated using standard AIPS procedures. The calibrated data were fitted by several circular Gaussian components using DIFMAP. The fitted components are labeled from C1 to C12, in inverse order of distance to the core. In Fig. \ref{VLBI} one can see a high resolution image of CTA\,102 observed at $15\,\mathrm{GHz}$ on 6th of January 2007. 

\begin{figure}[h!]
\resizebox{\hsize}{!}{\includegraphics[clip=true]{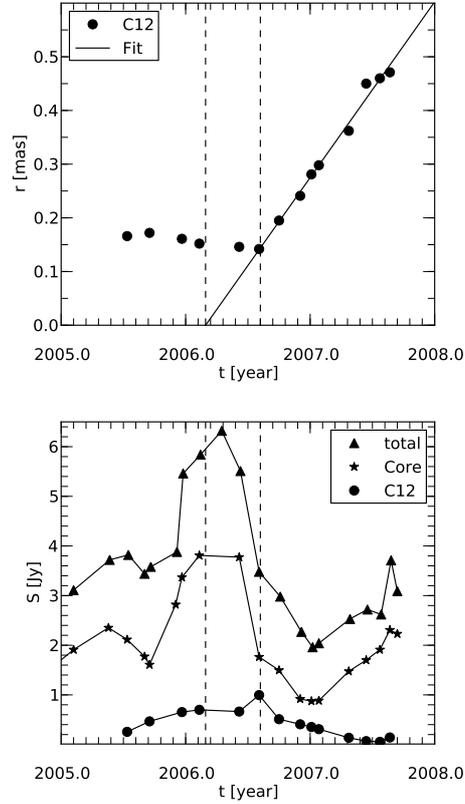}}
\caption{\footnotesize Top panel: Evolution of the separation from the VLBI core of component C12. Bottom panel: Flux density evolution of component C12, the VLBI core and the total flux density. The uncertainties in the flux densities are around 5\% and the positional errors are $\sim1/5$ of the restoring beam size (typically around 0.01\,$\mathrm{mas}$)}
\label{c12}
\end{figure}

Using the evolution of the flux density and displacement of the fitted components from the VLBI core we could identify the modeled features between the different epochs. This approach allows us to identify the feature labeled C12 as a possible candidate for a newly ejected component. The evolution of the flux density and position of component C12 are presented in Fig. \ref{c12}.

The separation from the core of component C12 remains constant until 2006.6, followed by a sharp acceleration and strong decrease in the observed flux density. \citet{Lis09b} derived an apparent speed of $8.5\pm 1.0\,\mathrm{c}$ and an ejection time of $t_\mathrm{ej}=2005.04\pm0.3\,\mathrm{yr}$ including the stationary phase before mid 2006 in the fit. For our analysis we separated the C12 trajectory into a stationary section, C12$_\mathrm{a}$, (before 2006.6)  and moving section, C12$_\mathrm{b}$ (after 2006.6). For C12$_\mathrm{b}$ we derived an apparent velocity of $17.3\pm 0.7\,\mathrm{c}$ and an ejection time of $t_\mathrm{ej}=2006.16\pm0.05\,\mathrm{yr}$.

\section{Discussion}
There is a flare in 15\,$\mathrm{GHz}$ observations of CTA\,102 around 2005.8 (see increase in total and core flux density in bottom panel of Fig.\,\ref{c12}). Furthermore, as already mentioned, the trajectory of component C12 is stationary until 2006.6. In the evolution of the flux density of this component one can see a global maximum around this time (see Fig. \ref{c12}).
\newline A possible scenario which explains the trajectory of C12 and the evolution of the flux density could be an interaction between traveling and standing shocks.
\newline The observed flare in CTA\,102 could be explained by the propagation of a relativistic shock wave. This wave is generated by pressure mismatches at the jet nozzle and during its way downstream it reaccelerates the underlying jet particles at the shock front. This interaction of the shock with the jet flow leads to an increased emissivity and can be observed with VLBI techniques \citep{Mar09}.
\newline The stationarity of component C12$_\mathrm{a}$ could be associated with a standing shock around $0.15\,\mathrm{mas}$ away from the core (see Fig. \ref{c12}). A standing shock is a stationary wave in non-pressure matched jets, created during the collimation of the jet \citep{Dal88,Fal91}. The position of such a shock remains stationary as long as the relative conditions of the jet and ambient medium remain unchanged.
\newline Using the trajectory of component C12$_\mathrm{b}$, we derived an ejection time of $t_\mathrm{ej}\,=\,2006.16\,\pm\,0.05\,\mathrm{yr}$. From this result it is not immediately obvious that one can associate the C12$_\mathrm{b}$ with a travelling shock wave which triggered the observed flare around 2005.8. This temporal discrepancy between the onset of the flare and the calculated ejection time of component C12$_\mathrm{b}$ could be explained in the following way: Due to the resolution limit of the 15\,$\mathrm{GHz}$ VLBI observations there is a lack of data less than $0.15\,\mathrm{mas}$ from the core. By using a linear interpolation based on the observed trajectory of component C12$_\mathrm{b}$, we implicitly assumed a constant velocity for this component and do not take acceleration into account \citep{Hom09}. If there is acceleration in the trajectory of C12$_\mathrm{b}$, this feature could have been ejected earlier than the calculated 2006.16. Therefore we could identify the C12$_\mathrm{b}$ trajectory with the interaction of a traveling shock wave with the underlying jet flow causing the 2005.8 flare.
\newline Such a moving shock wave would reach its peak flux density soon after ejection from the core, followed by a permanent flux density decrease due to, e.g., adiabatic expansion. In the case of CTA\,102 there is  a stationary feature around $0.15\,\mathrm{mas}$ from the core and therefore this effect can not be observed due to blending with the emission coming form the stationary feature (nearly constant flux density after 2006.2 of component C12). While the traveling shock is propagating downstream, it will collide with the stationary one after some time $t_{col}=2006.6$. During this collision the traveling shock will encounter a local increase in the particle density compared to the surrounding flow. The higher density at this position would then lead to an increase in the emissivity because of the shock front accelerating more particles. This process could explain the global flux density maximum of component C12 around 2006.6. Furthermore, the standing shock would be pulled away by the moving one and after some time the stationary feature should re-appear at the same position \citep{Mim09}.
\newline This scenario is in good agreement with the observed flux density evolution of the component C12. We could not detect the stationary section, C12$_\mathrm{a}$, after the collision of the two waves in 2006.6. A possible explanation could be the limited dynamic range of the VLBI observations or changes in the ambient medium configurations during the collision of the traveling shock wave. \newline To confirm our suggestion of a shock-shock interaction in the jet of CTA\,102 one should perform a spectral analysis on the available single-dish and multi-frequency VLBI observations. The overall behaviour in the turnover frequency turnover flux density plane is in agreement with the standard shock-in-jet model with a double hump feature. This feature could be explained by an increase in the Doppler factor or by a shock-shock interaction. The results of this analysis will be published elsewhere. Future research on the shock-shock interaction should also include relativistic (magneto) hydrodynamic (RMHD) simulations. These simulations could help us to better understand the process during the interaction of the traveling shock wave with the stationary one and could bridge the observational time gaps. Furthermore these calculations could provide an estimate for the time of re-appearance of the standing shock wave.

\section{Conclusions}
We presented the scenario of an interaction between traveling and standing shock as an explanation for the observed 2006 radio flare in CTA\,102. Furthermore this scenario could explain the trajectory of component C12, which could be associated with the flaring event.
\newline The combination of multi-frequency observations (single-dish as well as multi-frequency VLBI observations) and R(M)HD simulations will allow us to confirm, or not, the assumption of a shock-shock interaction in CTA\,102. The analysis of single-dish observations in the framework of the shock-in-jet model \citep{Mar85} will also provide estimates of the evolution of the magnetic field, Doppler factor and spectral index of the relativistic electron distribution (Fromm et al. 2010, in prep.).  

\begin{acknowledgements}
We thank R. W. Porcas and F. K. Schinzel for valuable comments and inspiring discussions.
\newline C.M. Fromm.  was supported for this research through a stipend from the International Max Planck Research School (IMPRS) for Astronomy and Astrophysics at the universities of Bonn and Cologne. 
\newline M. Perucho acknowledges support from a ``Juan de la Cierva'' contract of the Spanish
``Ministerio de Ciencia y Tecnolog\'{\i}a'', the Spanish ``Ministerio de Educaci\'on y Ciencia'' and the European Fund for Regional Development through grants AYA2007-67627-C03-01 and AYA2010-21322-C03-01 and Consolider-Ingenio 2010, ref. 20811. Part of this work was carried out when T. Savolainen was supported by Alexander von Humboldt foundation. 
\newline This research has made use of data from the MOJAVE database that is maintained by the MOJAVE team \citep{Lis09b}.
\end{acknowledgements}

\bibliographystyle{aa}

\end{document}